\documentclass[pra,twocolumn,showpacs,superscriptaddress,floatfix,amsmath,nofootinbib,amssymb]{revtex4}
\usepackage{dcolumn}
\usepackage{bm}
\usepackage{graphics}
\usepackage{epstopdf}
\usepackage{epsfig}
\usepackage{subfigure}
\def\b{\begin{equation}}
\def\e{\end{equation}}

\def\ket#1{|#1\rangle }

\pacs{03.67.Mn, 03.65.-w, 04.62.+v}
\begin{document}
\title{Entanglement of Dirac fields in an expanding spacetime}
\author{Ivette Fuentes\footnote{Previously known as Fuentes-Guridi and Fuentes-Schuller.}}
\address{School of Mathematical Sciences, University of Nottingham, Nottingham NG7 2RD,
United Kingdom}
\author{Robert B. Mann}
\affiliation{Department of Physics, University of Waterloo,
Waterloo, Ontario Canada N2L 3G1}
\author{Eduardo Mart\'{i}n-Mart\'{i}nez}%
 \affiliation{Instituto de F\'{i}sica Fundamental, CSIC, Serrano 113-B, 28006 Madrid, Spain}
\author{Shahpoor Moradi}
\address{Department of Physics, Razi University, Kermanshah, Iran}
\date{\today}

\begin{abstract}
We study the entanglement generated between Dirac modes in a 2-dimensional conformally flat Robertson-Walker universe.  We find radical qualitative differences between the bosonic and fermionic entanglement generated by the expansion. The particular way in which fermionic fields get entangled encodes more information about the underlying space-time than the bosonic case, thereby allowing us to reconstruct the parameters of the history of the expansion. This highlights the importance of bosonic/fermionic statistics to account for relativistic effects on the entanglement of quantum fields.

\end{abstract}

\maketitle
\section{Introduction}
The phenomenon of entanglement has been extensively studied in
non-relativistic settings. Much of the interest on this quantum
property has stemmed from its relevance in quantum information
theory. However, relatively little is known about relativistic
effects on entanglement
\cite{Czachor1,peresterno,Alsingtelep,TeraUeda2,ShiYu,Alicefalls,AlsingSchul,SchExpandingspace,Adeschul,LingHeZ,ManSchullBlack,PanBlackHoles,DH,Edu1,Steeg,Edu2,Shapoor,Ditta,Hu,DiracDiscord,Edu6} despite the fact that many of
the systems used in the implementation of quantum information
involve relativistic systems such as photons. The vast majority of
investigations on entanglement assume that the world is flat and
non-relativistic. Understanding entanglement in spacetime is
ultimately necessary because the world is fundamentally
relativistic. Moreover, entanglement  plays a prominent
role in black hole thermodynamics \cite{bombelli,Canent,Terashima,Emparan,levay,Cadoni,Hu2,NavarroSalas} and in the
information loss problem \cite{Maldacena,Preskill,Lloyd2,Ahn1,ManSchullBlack, schacross}.

Recently, there
has been increased interest in understanding entanglement and
quantum communication in black hole spacetimes
\cite{Xian,Pan2,Ahntropez,Edu6} and in using quantum information techniques
to address questions in gravity \cite{Ternada,Ternada2}. Studies on
relativistic entanglement show that conceptually important
qualitative differences to a non-relativistic treatment arise. For
instance, entanglement was found to be an observer-dependent
property that is degraded from the perspective of accelerated
observers moving in flat spacetime \cite{Alicefalls,AlsingSchul,Edu2,Adeschul,Villalba}.
These results suggest that entanglement in curved spacetime might
not be an invariant concept.

In this paper we study the creation of entanglement
between Dirac modes due to the expansion of a Robertson-Walker spacetime.
A general study of entanglement   in curved spacetime is problematic because particle states cannot always be defined in a meaningful way. However,
 it has been possible
 to learn about certain aspects of entanglement in curved spacetimes
that have asymptotically flat regions \cite{ball,schacross,ShiYu,TeraUeda2}. Such
studies show that entanglement can be created by the dynamics of the
 underlying spacetime \cite{ball,Steeg} as well as destroyed by the loss of
information in the presence of a spacetime horizon \cite{Alicefalls,schacross,Edu6}.

Such investigations not only deepen our understanding of entanglement but also offer the prospect of employing entanglement as a tool to learn about curved spacetime.  For example, the entanglement generated between bosonic modes due to the expansion of a model 2-dimensional universe was shown to contain information about its history \cite{ball}, affording the
possibility of deducing cosmological parameters of the underlying
spacetime from the entanglement. This novel way of obtaining
information about cosmological parameters could provide new insight into
the early universe  both theoretically (incorporating into cosmology entanglement as a purely quantum effect produced by gravitational interactions in an expanding universe) and experimentally (either by development of methods to measure entanglement between modes of the background fields or by measuring entanglement creation in  condensed matter analogs of expanding  space-time \cite{analog1,analog2}). Other interesting results show that entanglement plays a role
in the thermodynamic properties of Robertson-Walker type spacetimes
\cite{Lousto} and can in principle be used to distinguish between different spacetimes
 \cite{Steeg} and probe spacetime fluctuations \cite{dowling}.

Here we consider entanglement between modes of a Dirac field in
a  2-dimensional Robertson-Walker universe.  We find
that the entanglement generated by the expansion of the
universe for the same fixed conditions is lower than for the bosonic case \cite{ball}.
However we also find that fermionic entanglement  codifies more information about the underlying spacetime structure. These contrasts are commensurate with 
the flat spacetime case, in which entanglement in fermionic systems was found to
be more robust against acceleration than that in bosonic systems
\cite{AlsingSchul,Edu2}. In the limit of infinite acceleration fermionic entanglement remains finite due to statistical effects \cite{edu3,edu4} which resemble those  found in first quantization scenarios \cite{sta1}.

Our paper is organized as follows: In section \ref{sec2} we
revise the Dirac equation in a spatially flat $d$-dimensional Robertson-Walker
universe. Subsequently setting $d=2$, in section \ref{sec3} we calculate the entanglement entropy of two Dirac modes  and compare it with the bosonic case. In section \ref{sec4} we explain the origin of the entanglement peculiarities of the fermionic case, showing how it can give us more information about the parameters of the expansion. Conclusions are presented in section \ref{conclusions}.

\section{Dirac field in a $d$-dimensional Robertson-Walker universe}\label{sec2}

As we mentioned before, entanglement between modes of a quantum field in curved spacetime can be investigated in special cases where the spacetime has at least two asymptotically flat regions. Such is the case of the Robertson-Walker universe where spacetime is flat in the distant past and in the far future. In this section, following the work done by Bernard and Duncan \cite{BernardDuncan,dun1}, we find the state of a Dirac field in the far future that corresponds to a vacuum state in the remote past.

Consider a Dirac field $\psi$ with mass $m$ on a $d$-dimensional spatially flat Robertson-Walker spacetime with line element, \b
ds^2=C(\eta)(-d\eta^2+dx_idx^i).\e $x_i$ are the spacial coordinates and the temporal coordinate $\eta$ is called the conformal
time to distinguish it from the cosmological time $t$.
The metric is conformally flat, as are
all Robertson-Walker metrics.  The dynamics of the field is given by the covariant
form of the Dirac equation on a curved background,\b \label{eq:dirac}
\{i\gamma^{\mu}(\partial_{\mu}-\Gamma_{\mu})+m\}\psi=0, \e
where $\gamma^{\mu}$ are the curved Dirac-Pauli matrices and $\Gamma_{\mu}$ are spinorial
affine connections. The curved Dirac-Pauli matrices satisfy the condition, \b
\gamma^{\mu}\gamma^{\nu}+\gamma^{\nu}\gamma^{\mu}=2g^{\mu\nu}, \e  where $g^{\mu\nu}$ is the spacetime metric.
In the flat case where the metric is given by $\eta^{\alpha\beta}$, the constant special relativistic matrices are defined by, \b
\bar{\gamma}^{\alpha}\bar{\gamma}^{\beta}+\bar{\gamma}^{\beta}\bar{\gamma}^{\alpha}=2\eta^{\alpha\beta}
.\e The relation between curved and flat ${\gamma}$ matrices is given by $\gamma_{\mu}=e_{\mu}^{\;\;\alpha}\bar{\gamma}_{\alpha}$ where
$e_{\mu}^{\;\;\alpha}$ is the vierbein (tetrad) field satisfying
the relation $e_{\mu}^{\;\;\alpha}e_{\nu}^{\;\;\beta}\eta_{\alpha\beta}=g_{\mu\nu}$.

In order to find the solutions to the Dirac equation Eq.
(\ref{eq:dirac}) on this spacetime, we note that $C(\eta)$ is
independent of x. We exploit the resulting spatial translational
invariance and separate the solutions into 
\b 
\psi_k(\eta,x)
=e^{i\bm{k\cdot x}}C^{(1-d)/4}\left(\bar{\gamma} ^{0}\partial
_{\eta
}+i\bar{\bm{\gamma}}\cdot{{\bm k}}-mC^{1/2}\right)\phi_k(\eta)
,\e
where $k^2=|{{\bm k}}|^2= \sum_{i=1}^{d-1}k_i^2 $.
Inserting this into the Dirac equation, we obtain the following
coupled equations   \b \label{eq:couple} \left(\partial_{\eta
}^{2}+m^{2}C\pm
im\dot{C}C^{-1/2}+|{{\bm k}}|^{2}\right)\phi_k^{(\pm)}=0, \e
using the fact that the eigenvalues of $\bar{\gamma}^0$ are $\pm
1$. In order to quantize the field and express it in terms of
creation and annihilation operators, positive and negative frequency modes
must be identified. This cannot be done globally. However
positive and negative frequency modes can be identified in the far
past and future where the spacetime admits timelike killing vector
fields $\pm\partial/\partial \eta$. Provided $C(\eta)$ is constant
in the far past $\eta\rightarrow -\infty$ and
far future $\eta\rightarrow +\infty$, the asymptotic solutions of Eq.
(\ref{eq:couple}) will be $\phi^{(\pm)}_{in}\sim
e^{\pm i\omega_{in}\eta}$ and $\phi^{(\pm)}_{out}\sim e^{\pm
i\omega_{out}\eta}$  respectively, where 
 \begin{eqnarray}
\omega_{in}&=&(|{\bm k}|^2+\mu_{in}^2)^{1/2} \\
\omega_{out}&=&(|{\bm k}|^2+\mu_{out}^2)^{1/2} \nonumber\\
\mu_{in}&=&m\sqrt{C(-\infty)},\nonumber\\\mu_{out}&=&m\sqrt{C(+\infty)}. \nonumber\end{eqnarray}

The action of the Killing vector field on the asymptotic solutions allow us to identify $\phi^{(\mp)\ast}_{in}$ and $\phi^{(\mp)\ast}_{out}$ as negative frequency  solutions. The sign flip is due to the explicit factor $i$ in (\ref{eq:couple}).
A consequence of the linear transformation
properties of such functions is that
the Bogolubov transformations associated with the transformation between $in$ and $out$ solutions take the simple form \cite{dun1}
\b
\phi_{in}^{(\pm)}(k)= \alpha_{k}^{(\pm)}\phi^{(\pm)}_{out}(k)+\beta_{k}^{(\pm)}\phi^{(\mp)*}_{out}(k)
,\e
where $\alpha^{\pm}_k$ and $\beta^{\pm}_k$ are Bogoliubov coefficients.

The curved-space spinor solutions of the Dirac equation are defined by (with corresponding $U_{out}$, $V_{out}$ and $K_{out}$),
\begin{eqnarray}
\nonumber U_{in}(\bm{k},\lambda;x,\eta)\!\!&\equiv&\!\! K_{in}(k)[C(\eta)]^{(1-d)/4}\Big[-i\partial_\eta+i{\bm{k}}\cdot\bar{\gamma}\\*
&&\!\!-m\sqrt{C(\eta)}\Big]\phi_{k}^{in(-)(\eta)}e^{i\bm{k}\cdot
{\bm{x}}}u(0,\lambda)\nonumber\\
\nonumber V_{in}(\bm{k},\lambda;x,\eta)\!\!&\equiv&\!\!
K_{in}(k)[C(\eta)]^{(1-d)/4}\Big[i\partial_\eta-i{\bm{k}}\cdot\bar{\gamma}\\*
&&\!\!-m\sqrt{C(\eta)}\Big]\phi_{k}^{in(+)\ast(\eta)}e^{-i\bm{k}\cdot
{\bm{x}}}v(0,\lambda),\nonumber\\
\end{eqnarray}
where $K_{in}\equiv-(1/|k|)((\omega_{in}-\mu_{in})/2\mu_{in})^{1/2}$ and $u(0,\lambda)$, $v(0,\lambda)$ are flat space spinors satisfying,
\begin{eqnarray}
\gamma^0u(0,\lambda)&=&-iu(0,\lambda),\nonumber\\
\gamma^0v(0,\lambda)&=& iv(0,\lambda),\nonumber
\end{eqnarray}
for $1\leq\lambda\leq2^{d/2-1}$. The field in the ``in" region can
then be expanded as,

\begin{eqnarray}
\nonumber \psi(x)&=&\frac{1}{\sqrt{(2\pi)^{1-d}}}\int
d^{d-1}k\left[\frac{\mu_{in}}{\omega_{in}}\right]\sum_{\lambda=1}^{d/2-1}
\big[a_{in}({\bm k},\lambda)\\*
&&U_{in}({\bm k},\lambda;\bm{x},\eta)+
 b^{\dag}_{in}({\bm k},\lambda)V_{in}({\bm k},\lambda;\bm{x},\eta)\big]
\end{eqnarray} 
with a similar expression for the ``out" region.  The $in$
and $out$ creation and annihilation operators for particles and
anti-particles obey the usual anticommutation relations. Using the
Bogoliubov transformation one can expand the $out$  operators in
terms of $in$ operators 
\begin{eqnarray}
a_{out}(\bm k,\lambda)&=&\nonumber\left(\frac{\mu_{in}\omega_{out}}{\omega_{in}\mu_{out}}\right)^{\frac{1}{2}}
\frac{K_{in}}{K_{out}}\Bigg(\alpha_{k}^{(-)}a_{in}(\bm k,\lambda)\\*
&+&\beta_{k}^{(-)\ast}\sum_{\lambda'}
X_{\lambda\lambda'}(-{\bm k})b^{\dag}_{in}(-{\bm k},\lambda')\Bigg),\\*
\nonumber b_{out}(\bm k,\lambda)&=&\left(\frac{\mu_{in}\omega_{out}}{\omega_{in}\mu_{out}}\right)^{\frac{1}{2}}
\frac{K_{in}}{K_{out}}\Bigg(\alpha_{k}^{(-)}b_{in}(\bm k,\lambda)\\*
&+&\beta_{k}^{(-)\ast}\sum_{\lambda'}
X_{\lambda\lambda'}(-{\bm k})a^{\dag}_{in}(-{\bm k},\lambda')\Bigg),
\end{eqnarray}
where \b
X_{\lambda\lambda'}(-{\bm k})=-2\mu_{out}^2K^2_{out}\bar{u}_{out}(-{\bm k},\lambda')v(0,\lambda).
\e and \b
K_{in/out}=\frac{1}{|k|}\left(\frac{\omega_{in/out}(k)-\mu_{in/out}}{\mu_{in/out}}\right)^{1/2}
.\e This yields the following relationship between Bogoliubov
coefficients, 
\b
\left|\alpha^{(-)}_{k}\right|^2\!\!\!-2\mu_{out}^2K_{out}^2\left(1-\frac{\omega_{out}}{\mu_{out}}\right)\!\!\left|\beta_{k}^{(-)}\right|^2
\!\!\!\!=\!\frac{\mu_{out}}{\mu_{in}}\frac{\omega_{in}}{\omega_{out}}\!\left(\frac{K_{out}}{K_{in}}\right)^2\!\!\!\!
.\e We consider the special solvable case presented in \cite{dun1}
$C(\eta)=(1+\epsilon(1+\tanh\rho\eta))^2$, where $\epsilon,\rho $
are positive real parameters controlling the total volume and
rapidity of the expansion, respectively. In this case the
solutions of the Dirac equation  that  in remote past reduce to
positive frequency modes are,
\begin{eqnarray}
\nonumber\phi^{(\pm)}_{in}&=&\exp \left( -i\omega _{+}\eta-\frac{
i\omega _{-}}{\rho }\ln [2\cosh \rho \eta ]\right)\\*
&&\nonumber\!\!\!\!\!\!\!\!\!\!\!\!\!\!\!\!\!\!\!\!\!\!\! \times F_{1}\!\left(\!
1\!+\!\frac{
i(\omega _{-}\!\pm m\epsilon)}{\rho },\frac{i(\omega
_{-}\!\mp m\epsilon)}{\rho },1\!-\!\frac{
i\omega _{in}}{\rho },\frac{1\!+\!\tanh (\rho \eta )}{2}\!\right)\!\!,
\end{eqnarray}
where $F_{1}$ is the ordinary hypergeometric function. Similarly, one may
find a complete set of modes of the field that behaving as
positive and negative frequency modes in the far future, 
\begin{eqnarray}
\nonumber \phi^{(\pm)}_{out}&=&\exp \left( -i\omega _{+}\eta-\frac{%
i\omega _{-}}{\rho }\ln [2\cosh \rho \eta ]\right)\\*
&&\nonumber\!\!\!\!\!\!\!\!\!\!\!\!\!\!\!\!\!\!\!\!\!\!\!\! \times F_{1}\!\left(\!
1\!+\!\frac{
i(\omega _{-}\!\pm m\epsilon)}{\rho },\frac{i(\omega
_{-}\!\mp m\epsilon)}{\rho },\!1\!+\!\frac{
i\omega _{out}}{\rho },\frac{1\!-\!\tanh (\rho \eta )}{2}\!\right)\!\!,
\end{eqnarray}
 where $\omega _{\pm }=(\omega _{out}\pm \omega _{in})/2 $. The spacetime
obtained by considering this special form of $C(\eta)$ was
introduced by Duncan \cite{dun1}. It is easy to see that it
corresponds to a Minkowskian spacetime in the far future and past,
i.e., $ C\rightarrow (1+2\epsilon)^2$ in the $out$ region and
$C\rightarrow 1 $ at the $in$ region.

If we define $|\gamma^{-}|^2 \equiv \left|\beta^{(-)}_{k}/\alpha^{(-)}_{k}\right|^2$, for this spacetime we get that 
\begin{eqnarray}
\nonumber\left|\gamma^{-}\right|^2 &=&\frac{(\omega_- +m\epsilon)(\omega_+ +
m\epsilon)}{( \omega_-
- m\epsilon)(\omega_+ -m\epsilon)}\\
&\times&\!\!\frac{\sinh\left[\frac{\pi}{\rho} (\omega_- -
m\epsilon)\right]\sinh\left[\frac{\pi}{\rho} (\omega_- + m\epsilon)\right]}{\sinh\left[\frac{\pi}{\rho}(\omega_+
+m\epsilon)\right]\sinh\left[\frac{\pi}{\rho}(\omega_+ -m\epsilon)\right]}\label{gamma-F}
\end{eqnarray}

An analogous procedure can be followed for scalar fields \cite{ball}. The time dependent
Klein-Gordon equation in this spacetime is given by \b
\left(\partial_{\eta }^{2}+k^2+C(\eta) m^2\right)\chi_k(\eta)=0.
\e After some algebra, the solutions of Klein-Gordon equation
behaving as positive frequency modes as $\eta \rightarrow -\infty
(t\rightarrow -\infty )$, are found to be
\begin{eqnarray}
\chi_{in}(\eta)&=&\exp\left(-i\omega_+\eta-
 \nonumber \frac{i\omega_-}{\rho}\ln[2\cosh\rho\eta]\right)\\*
&&\nonumber\!\!\!\!\!\!\!\!\!\!\!\!\!\!\!\!\!\!\!\!\!\!\!\!\!\!\!\!\times F\!\left(\frac{1}{2}\!-\!\frac{i\bar{\omega}}{2\rho}\!+\!\frac{i\omega_-}{\rho},
\frac{1}{2}\!+\!\frac{i\bar{\omega}}{2\rho}\!+\!\frac{i\omega_-}{\rho},
1\!-\!i\frac{\omega_{in}}{\rho},\frac{1\!+\!\tanh(\rho\eta)}{2}\!\right)\!.
\end{eqnarray}
Similarly we have
\begin{eqnarray}
\chi_{out}(\eta)&=&\exp\left(-i\omega_+\eta-
 \frac{i\omega_-}{\rho}\ln[2\cosh\rho\eta]\right)\nonumber\\*
&&\nonumber\!\!\!\!\!\!\!\!\!\!\!\!\!\!\!\!\!\!\!\!\!\!\!\!\!\!\!\!\!\!\times F\!\left(\frac{1}{2}\!-\!\frac{i\bar{\omega}}{2\rho}\!+\!\frac{i\omega_-}{\rho},
\frac{1}{2}\!+\!\frac{i\bar{\omega}}{2\rho}\!+\!\frac{i\omega_-}{\rho},
1\!-\!i\frac{\omega_{out}}{\rho},\frac{1\!-\!\tanh(\rho\eta)}{2}\!\right)\!.
\end{eqnarray}
 where $\bar{\omega}=(m^2(2\epsilon+1)^2-\rho^2)^{1/2}$. Computing the quotient of the Bogoliubov coefficients for this bosonic case,  we find
 \b
|\gamma_B^{-}|^2=\frac{\cosh\frac{\pi}{\rho}\bar{\omega}
+\cosh\frac{2\pi}{\rho}\omega_-}
{\cosh\frac{\pi}{\rho}\bar{\omega}
+\cosh\frac{2\pi}{\rho}\omega_+}.  \label{gammaB}
\e

\section{Entanglement generated due to the expansion of the universe}\label{sec3}

It is then possible to find the state in the far future that
corresponds to the vacuum state in the far past. By doing that we will show that the vacuum state of the field in the asymptotic past evolves to an entangled state in the asymptotic future. The entanglement generated by the expansion codifies information about the parameters of the expansion, this information is more easily obtained from fermionic fields than bosonic, as we will show below.

Since we want to study fundamental behaviour we will consider the 2-dimensional case, which has all the fundamental features of the higher dimensional settings. 
 
 Using the relationship between particle operators in asymptotic times,
 \begin{equation}
b_{in}(k)=\left[{\alpha^{-}_{k}}^{\ast}b_{out}(k)+{\beta^{-}_{k}}^{\ast}%
\chi(k)a_{out}^{\dagger}(-k)\right],
\end{equation}
 we can obtain the asymptotically past vacuum state in terms of the asymptotically future Fock basis. Demanding that
$b_{in}(\bar{k},\lambda)|0\rangle_{in}=0$ we can find the ``in"
vacuum in terms of the ``out" modes. Due to the form of the Bogoliubov transformations the ``in" vacuum must be of the form 
$$
|0\rangle_{in}=\prod_{k}(A_{0}|0
\rangle_{out}+A_{1}|1_{k} 1_{-k}\rangle_{out} )
$$ 
where  to compress notation $\ket{1_{-k}}$ represents an antiparticle mode with momentum $-k$ and $\ket{1_k}$ a particle mode with momentum $k$. Here we wrote
the state for each frequency in the Schmidt decomposition. Since
different $k$ do not mix it is enough to consider only one
frequency.  Imposing $b_{in}(\bar{k})|0\rangle_{in}
=0$ we obtain the following condition on the vacuum coefficients
\begin{equation}
{\alpha^{-}_{k}}^*A_{1}|1_{-k}\rangle+
{\beta^{-}_{k}}^*\chi(\bar{k} )A_{0}|1_{-k}\rangle=0
\end{equation}
giving
\begin{eqnarray}
A_{1}=-\frac{ {\beta^{-}_{k}}^*}{{\alpha^{-}_{k}}^*}\chi(\bar{k}%
)A_{0}=-\gamma^{-\ast}\chi(\bar{k})A_{0} \end{eqnarray} where
\b\gamma^{-\ast}({k})=\frac{
{\beta^{-}_{k}}^*}{{\alpha^{-}_{k}}^*}\e From the vacuum
normalization,
\begin{eqnarray}
1&=&_{in}\langle 0|0\rangle_{in}=|A_{0}|^{2}(1+|\gamma^{-}({k})\chi(\bar{k}%
)|^{2}).
\end{eqnarray}
Therefore,  the vacuum state 
\begin{equation}
|0\rangle_{in}=\prod_{k}\frac{|0\rangle_{out}-\gamma^{-\ast}({k})\chi(\bar{k})|1_{k}
1_{-k}\rangle_{out} }{\sqrt{1+|\gamma^{-}({k})\chi(\bar{k})|^{2}}}
 \label{vacin}
\end{equation}
is an entangled state of particle modes and antiparticle modes with opposite momenta.

Since the state is pure, the entanglement is quantified by the
von-Neumann entropy given by
$S(\rho_k)=\mathrm{Tr}(\rho_k\log_2\rho_k)$ where $\rho_k$ is the
reduced density matrix of the state for mode $k$.  Tracing over the antiparticle 
modes with momentum $-k$ (or alternatively, particle modes with momentum $k$)  we obtain 
\b
\rho_{k}=\frac{1}{(1+|\gamma_k^{-}\chi(\bar{k})|^{2})}(|0\rangle%
\langle0|+|\gamma_k^{-\ast}({k})\chi(\bar{k})|^{2}|1_{k}
\rangle\langle1_{k}|). \e The von Neumann entropy of this state is
simply
\b
S(\rho_{k})=
\log(1+|\gamma_k^{-}\chi (\bar{k})|^{2})
-\frac{|\gamma_k^{-\ast}\chi (\bar{k})|^{2}\log(|\gamma_k^{-\ast}\chi (\bar{k})|^{2})}{(1+|\gamma_k^{-}\chi (%
\bar{k})|^{2})}
\label{Entropy} \e Using the following identity
\b
|\chi_{\lambda\lambda}|^{2}=
2\mu_{out}K^{2}_{out}(\mu_{out}-\omega_{out})=
 \left[\frac{%
\mu_{out}}{|k|}\left(1-\frac{\omega_{out}}{\mu_{out}}\right)\right]^{2} \e we
obtain the entanglement entropy
\begin{eqnarray}
\nonumber S(\rho_{k})
&=& \log\left(1+\frac{\mu^{2}_{out}}{|k|^{2}}\left(1-\frac{\omega_{out}}{%
\mu_{out}}\right)^{2}|\gamma_k^{-}|^{2}\right)\\*
\nonumber&-&\frac{\frac{\mu^{2}_{out}}{|k|^{2}}\left(1-\frac{\omega_{out}}{\mu_{out}}%
\right)^{2}|\gamma_k^{-\ast} |^{2}}{(1+\frac{\mu^{2}_{out}}{|k|^{2}}%
\left(1-\frac{\omega_{out}}{\mu_{out}}\right)^{2}|\gamma_k^{-}
|^{2})}\\
&\times&\log\left(\frac{\mu^{2}_{out}}{|k|^{2}}\left(1-\frac{\omega_{out}}{\mu_{out}%
}\right)^{2}|\gamma_k^{-\ast}(\bar{k})|^{2}\right)  \label{Ent2}
\end{eqnarray}

Using (\ref{Ent2}) we find that the fermionic entanglement is
\begin{equation}
S_F= \log\left(\frac{1+|\gamma_F^{-}|^{2}}{|\gamma_F^{-}|^{
\frac{2|\gamma_F^{-}|^{2}}{|\gamma_F^{-} |^{2}+1}}}\right)
\label{SF}
\end{equation}
where $|\gamma_F| = |\gamma_k^{-}\chi (\bar{k})| $. Note that for massless fields
($m=0$)  the entanglement vanishes since $\omega_-=0$ and $\gamma^-$=0.
Comparing  our result to the bosonic
case studied in \cite{ball} we find
\begin{equation}
S_B=\log\left(\frac{|\gamma_B^{-}|^{%
\frac{2|\gamma_B^{-}|^{2}}{|\gamma_B^{-}
|^{2}-1}}}{1-|\gamma_B|^{2}}\right) \label{Eq:entang}
\end{equation}
where the expression for $\gamma_B$ in (\ref{gammaB}) differs from that in ref.  \cite{ball} due to the different scale factor used.

The difference between the bosonic and fermionic cases means that the response of
entanglement  to the dynamics of the expansion of the universe depends on the nature of the quantum field. We see from \eqref{vacin} that  each fermionic field mode is always in a qubit state (the exclusion principle imposes a dimension-2 Hilbert space for the partial state). However in the bosonic case \cite{ball} the Hilbert space for each mode is of infinite dimension, as every occupation number state of the $out$ Fock basis participates in the $in$ vacuum.
In both cases the entanglement increases monotonically with the expansion rate $\rho$ and the total volume expansion parameter $\epsilon$. It is possible to find analytically the asymptotic values that both fermionic and bosonic entanglement reach at infinity. For example, when $k=m=\rho=1$ we find that as $\epsilon\rightarrow\infty$
\b
\gamma_B^{-} \to e^{-\pi\sqrt{2}}  \qquad  \gamma_F \to e^{-\pi\sqrt{2}} \frac{e^{\pi\sqrt{2}}-e^{\pi}}{e^{\pi\sqrt{2}+1}-1}
\e
respectively yielding
\b
 S_{B}(\epsilon\rightarrow\infty)\approx 0.0913 \qquad    S_{F}(\epsilon\rightarrow\infty)\approx 0.0048
 \e

The entanglement entropy is bounded by $S_E < \log_2 N$ where $N$ is the Hilbert space dimension of the partial state. The  fermionic upper limit $S_E=1$ corresponds to a maximally entangled state. For bosons the unbounded dimension of the Hilbert space implies the entropy of entanglement is not bounded by  unity  \cite{ball}. However this distinction does not guarantee that we can extract more information from bosons as we shall now demonstrate.

\section{Fermionic entanglement and the expansion of the universe}\label{sec4}

\begin{figure}[h]
\begin{center}
\includegraphics[width=.50\textwidth]{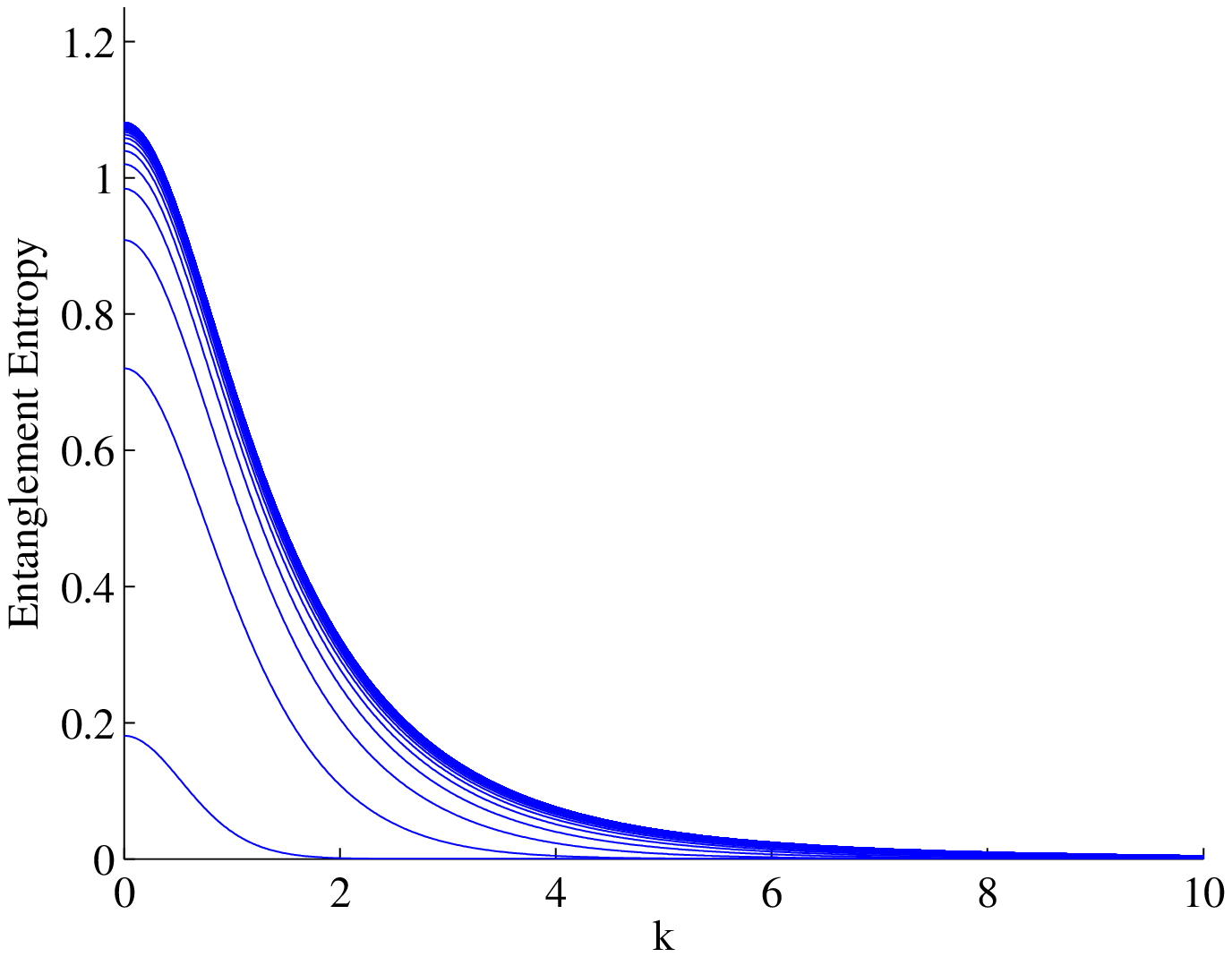}
\includegraphics[width=.50\textwidth]{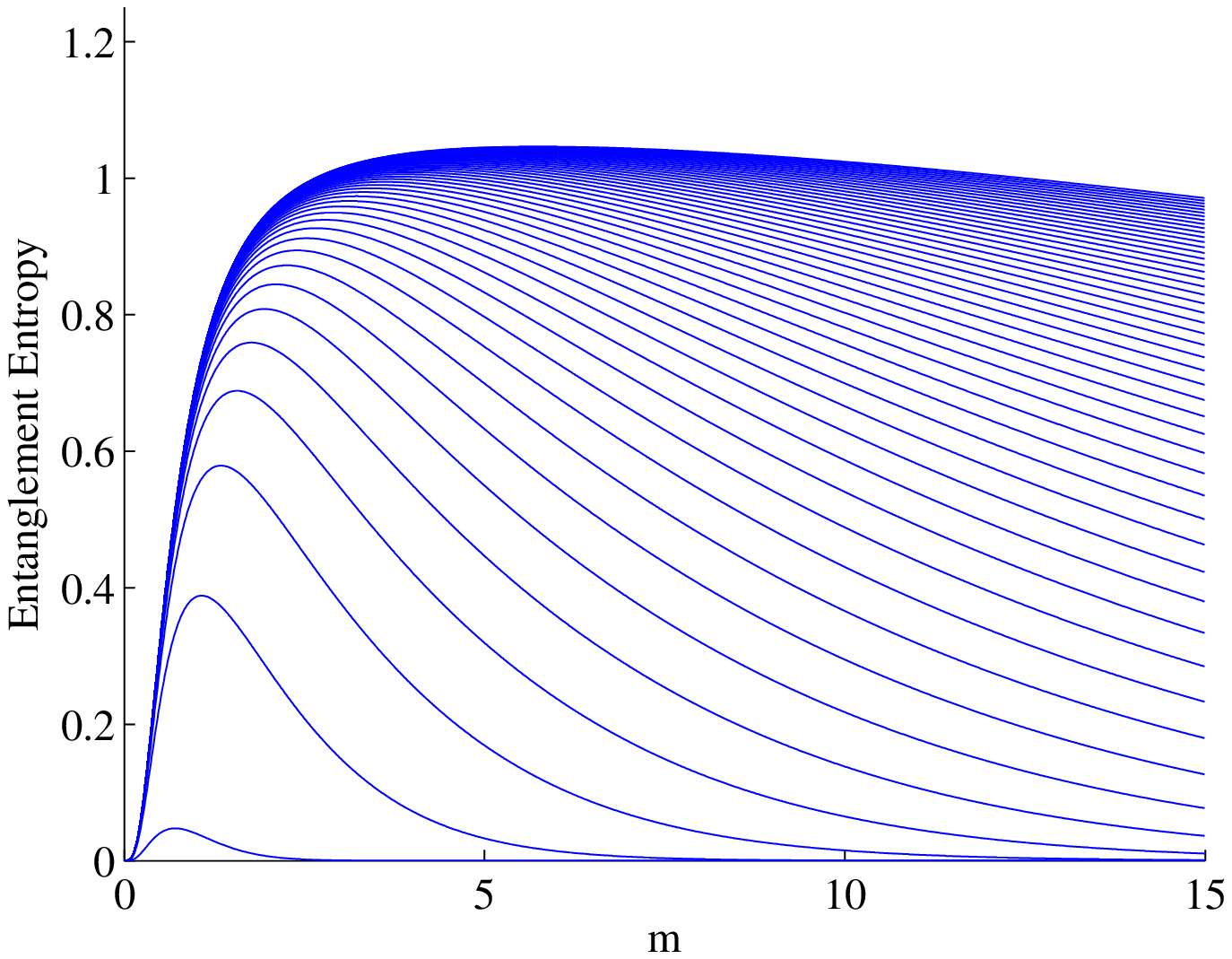}
\end{center}
\caption{Bosonic field: $S_E$ for a fixed mass $m=1$ as a funciton of $|\bm k|$ (up) and for a fixed $|\bm k|=1$ as a function of $m$ (down) for different rapidities $\rho=1,\dots,100$. An asymptotic regime is reached when $\rho\rightarrow\infty$. $\epsilon$ is fixed $\epsilon=1$}
\label{bosons}
\end{figure}

\begin{figure}[h]
\begin{center}
\includegraphics[width=.50\textwidth]{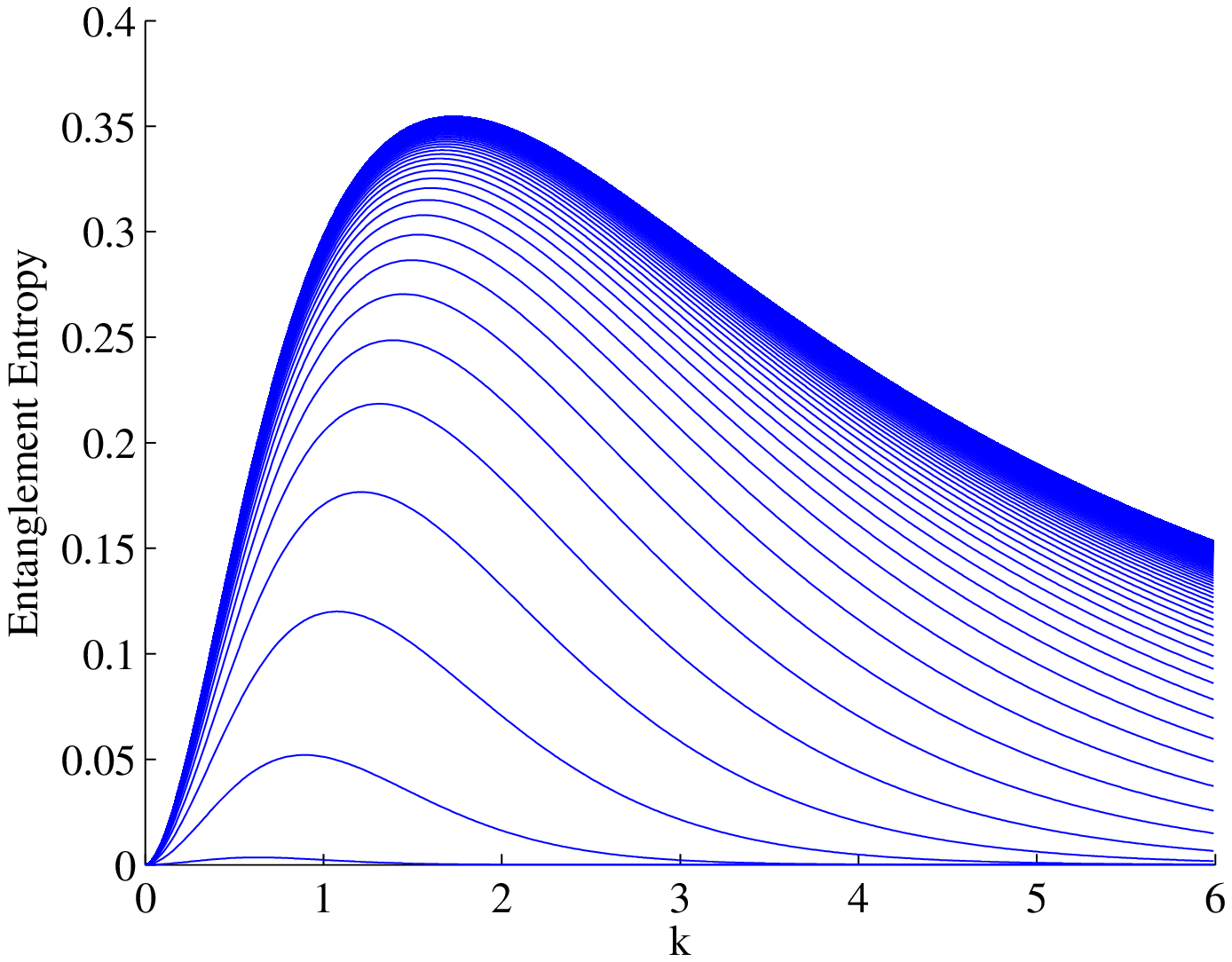}
\includegraphics[width=.50\textwidth]{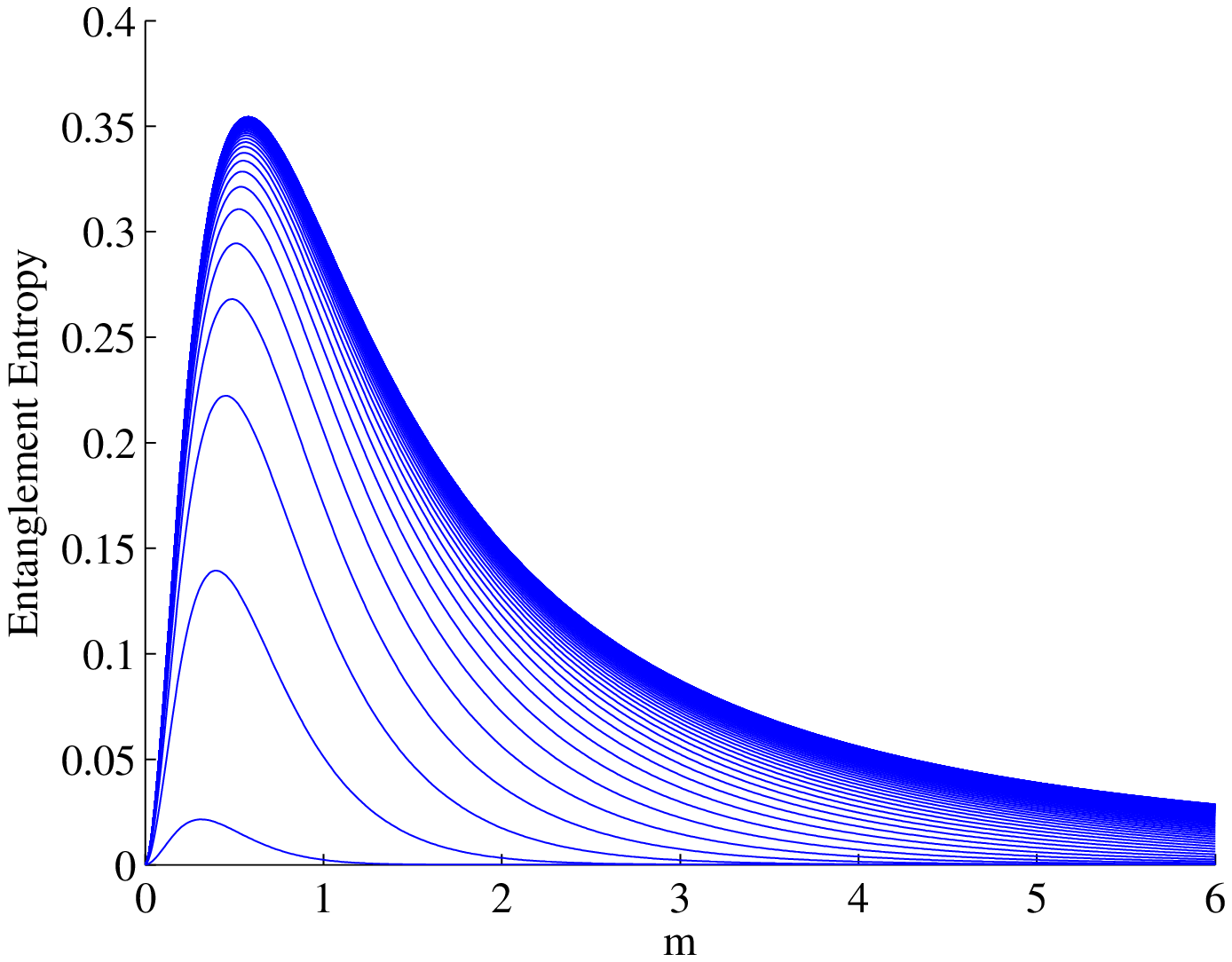}
\end{center}
\caption{ Fermionic field: $S_E$ for a fixed mass $m=1$ as a funciton of $|\bm k|$ (up) and for a fixed $|\bm k|=1$ as a function of $m$ (down) for different rapidities $\rho=1,\dots,100$. The maximum shifts to the right reaching an asymptotic value as $\rho\rightarrow\infty$. $\epsilon$ is fixed $\epsilon=1$. The behaviour as $|\bm k|$ varies is radically different from the bosonic case.}
\label{peaked}
\end{figure}

As seen in figures \ref{bosons} and \ref{peaked} the entanglement behaviour is completely different for bosons (fig. \ref{bosons}) and fermions (fig. \ref{peaked}). Although the behaviour as the mass of the field varies seems qualitatively similar, the variation with the frequency of the mode is completely different. 

The entanglement dependence on $|\bm k|$ for bosons is monotonically decreasing whereas for fermions, the global space-time structure `selects' one value of $|\bm k|$ for which the expansion of the space-time generates a larger amount of entanglement (peak in figure \ref{peaked}). We shall see that this selection of a privileged mode is sensitive to the expansion parameters. This may be related to the fermionic nature of the field insofar as the exclusion principle impedes entanglement for too small $|\bm k|$.

Regardless of its origin, we can take advantage of this special behaviour for fermionic fields to use the expansion-generated entanglement to engineer a method to obtain information about the underlying space-time more efficiently than for bosons. 

\subsection{Using fermionic fields to extract information from the ST structure}

\begin{figure}[h]
\begin{center}
\includegraphics[width=.50\textwidth]{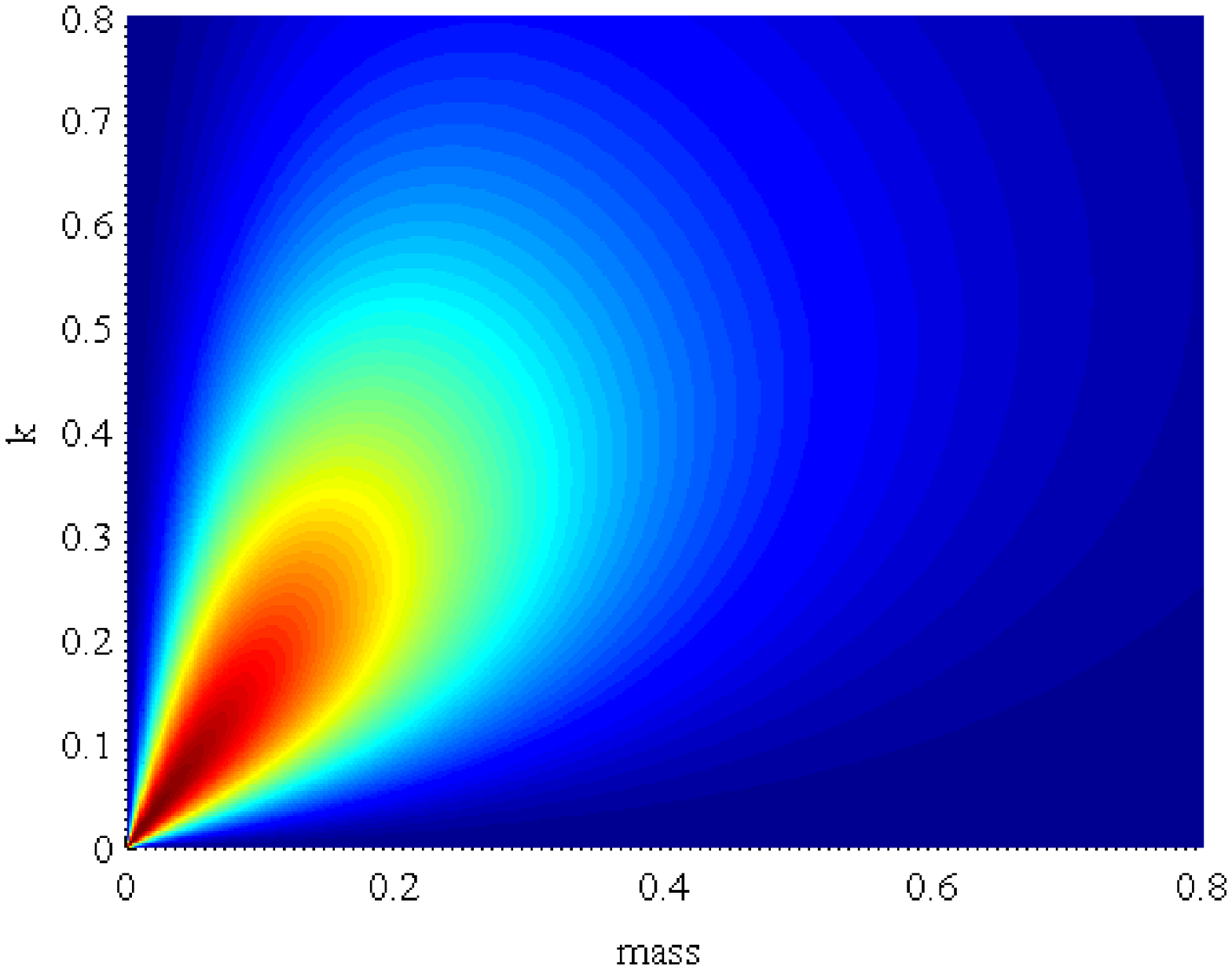}
\includegraphics[width=.50\textwidth]{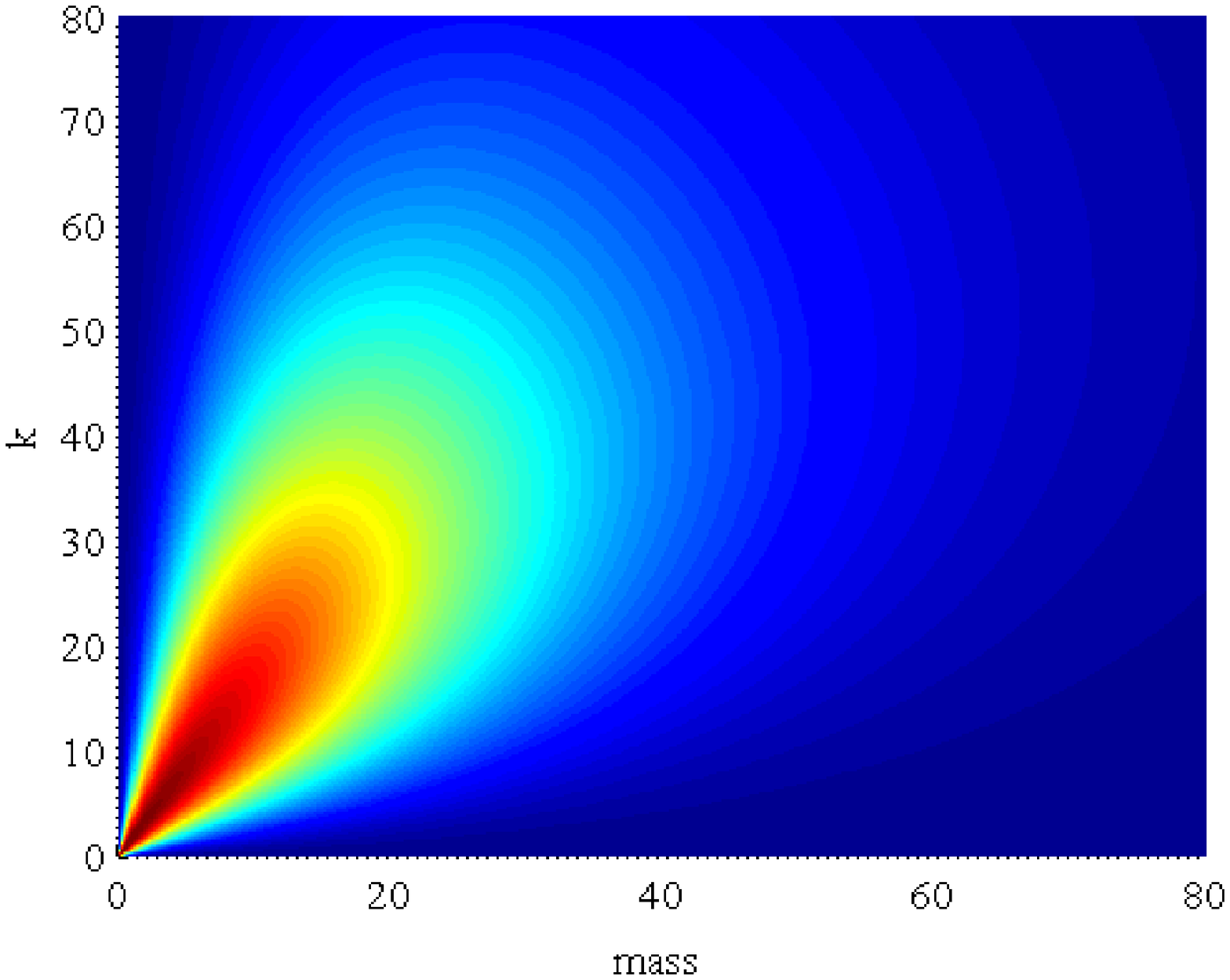}
\end{center}
\caption{(Color online) $S_E(m,|\bm k|)$ for $\epsilon=1$, $\rho=1$ (up) $\rho=100$ (down). Red color $\equiv$ higher $S_E(m,|\bm k|)$}
\label{fig3d2}
\end{figure}

Doing a conjoint analysis of the mass and momentum dependence of the entropy we can exploit the characteristic peak that $S_E(m,|\bm k|)$ presents for fermionic fields to obtain information from the underlying structure of the space-time better than we can do with a bosonic field. Let us first show both dependences simultaneously. Figure \ref{fig3d2} shows the entropy of entanglement as a function of our field parameters ($|\bm k|$ and mass)  for different values of the rapidity.

\begin{figure}[h]
\begin{center}
\includegraphics[width=.50\textwidth]{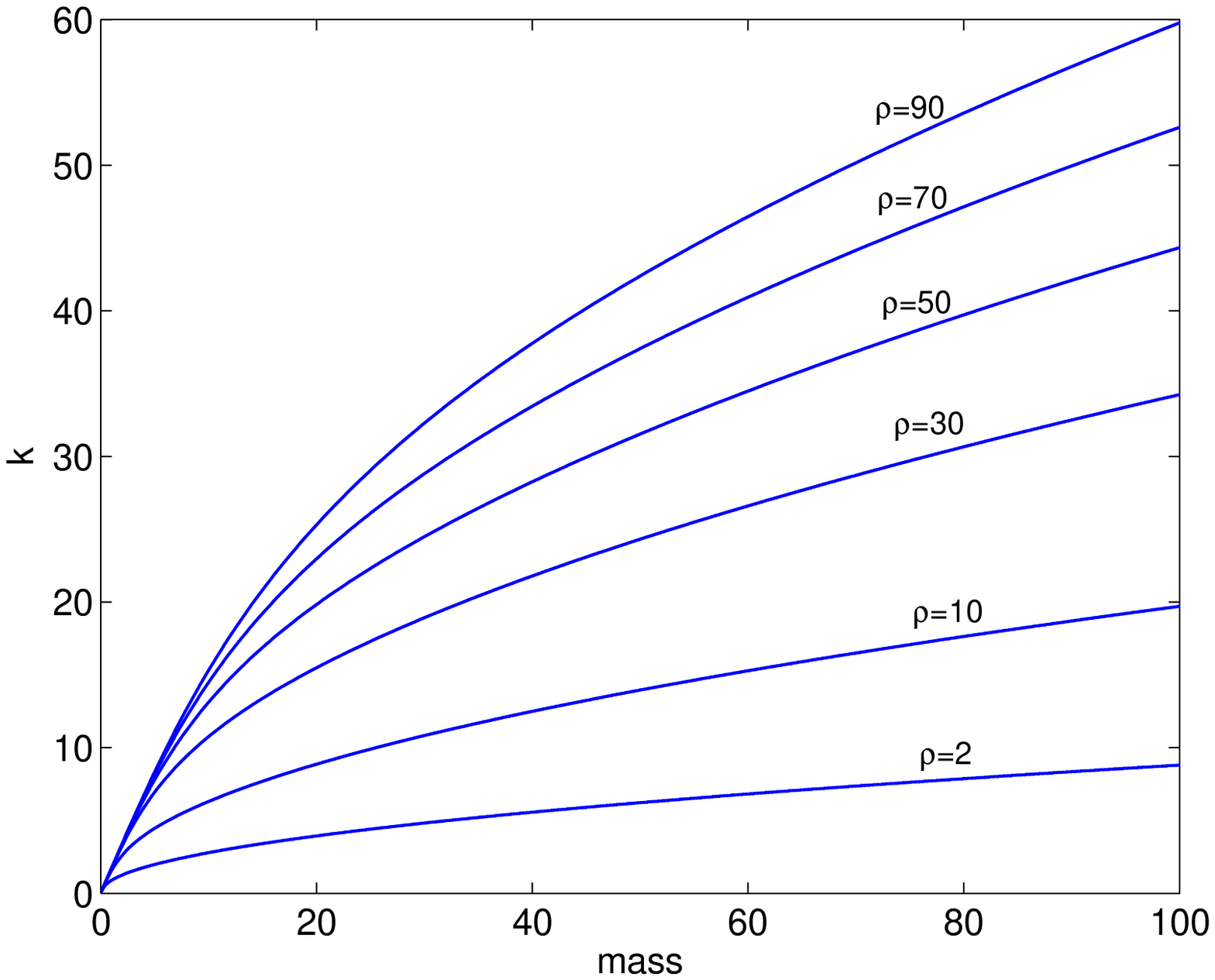}
\includegraphics[width=.50\textwidth]{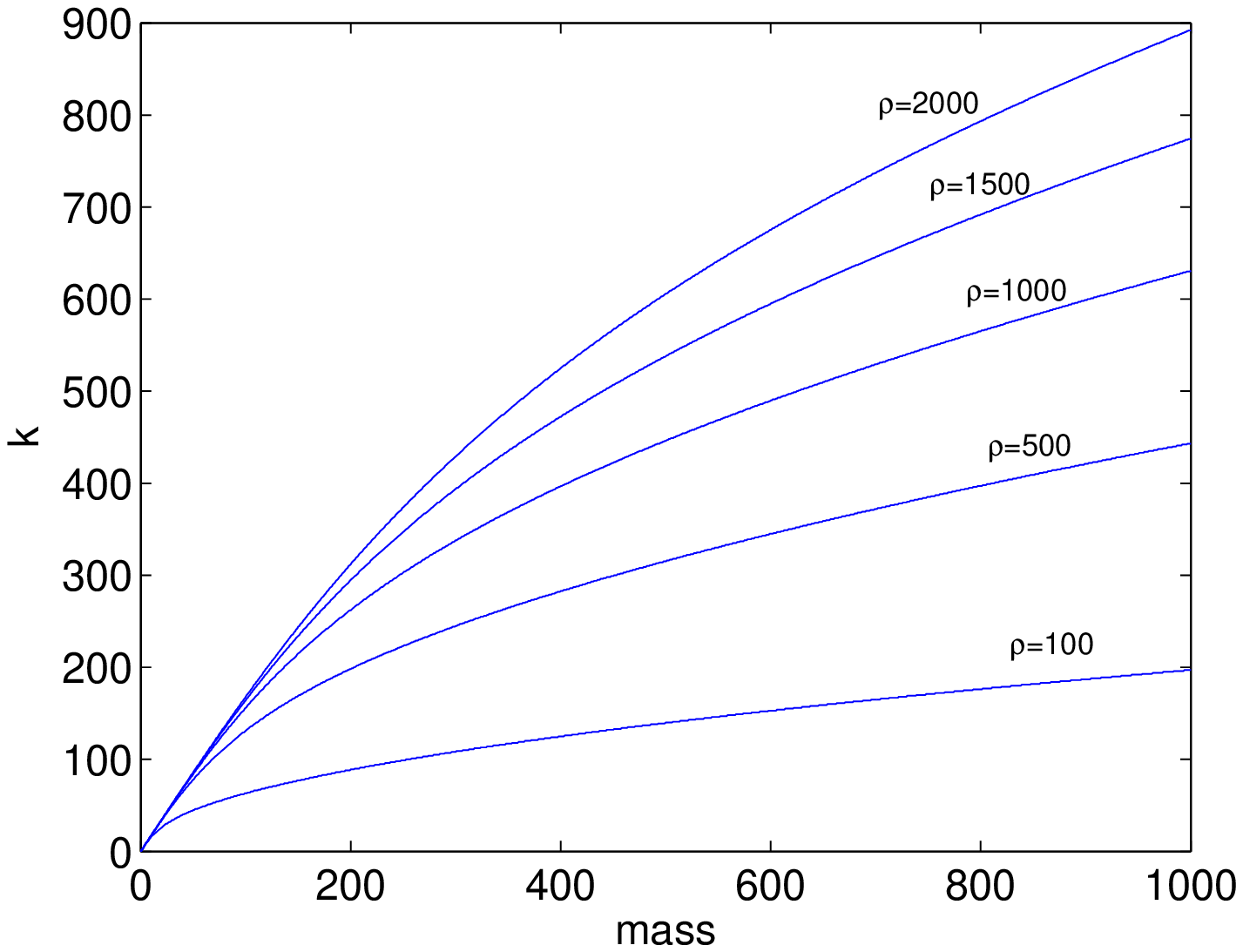}
\end{center}
\caption{$\epsilon=1$, optimal $|\bm k|$ curves (maximum entanglement mode) as a function of the field mass for $\rho=10,\dots,2000$}
\label{Fig8}
\end{figure}

We see from figure \ref{fig3d2} that there is no saturation as $\rho\rightarrow\infty$. Instead, as $\rho$ is increased the plot is just rescaled. This is crucial in order to be able to trace back the metric parameters from  entanglement creation.  
We also see from figure \ref{fig3d2}  that, for a given  field mass, there is an optimal value of $|\bm k|$ that maximises the entropy. In figure \ref{Fig8} we represent this optimal $|\bm k|$ as a function of the mass for different values of $\rho$, showing how the mode which get most entangled as a result of the spacetime expansion changes with the mass field for different rapidities.

From the figure we can readily notice two important features
\begin{itemize}
\item The optimal $|\bm k|$ curve is very sensitive to $\rho$ variations and  there is no saturation (no accumulation of these lines) as $\rho$ is increased.
\item There is always a field mass for which the optimal $|\bm k|$ clearly distinguishes arbitrarily large values of $\rho$.
\end{itemize}

\begin{figure}[h]
\begin{center}
\includegraphics[width=.50\textwidth]{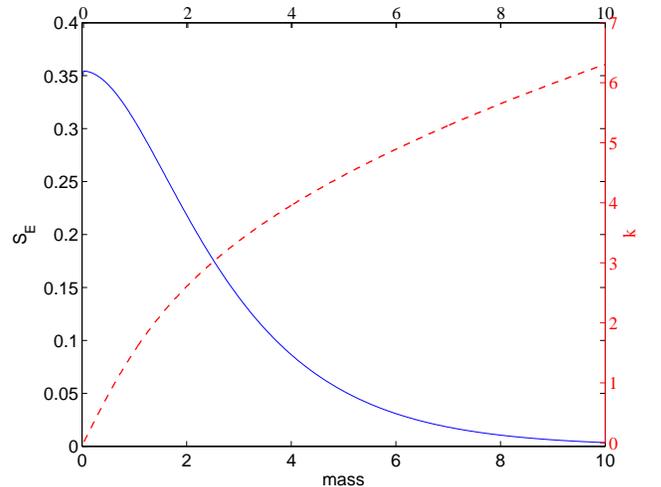}
\includegraphics[width=.50\textwidth]{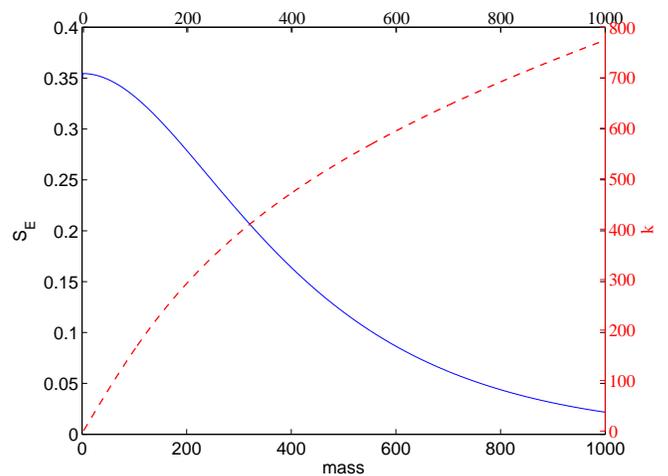}
\end{center}
\caption{(Color online) $S_E$ (blue continuous line) and $k$ (red dashed line) in the line of optimal $k$ as a function of mass for $\rho=10$ (up) and $\rho=1500$ (down). $\epsilon$ is fixed $\epsilon=1$}
\label{Fig10}
\end{figure}

In figure \ref{Fig10} we can see a consequence of the re-scaling (instead of saturation) of $S_E(m,|\bm k|)$ when $\rho$ varies. In this figure we show simultaneously the entropy in the optimal curve and the value of the optimal $|\bm k|$ as a function of the mass of the field for two different values of $\rho$, showing that if $\rho$ results to be very large,   entanglement decays more slowly for higher masses. 

\begin{figure}[h]
\begin{center}
\includegraphics[width=.50\textwidth]{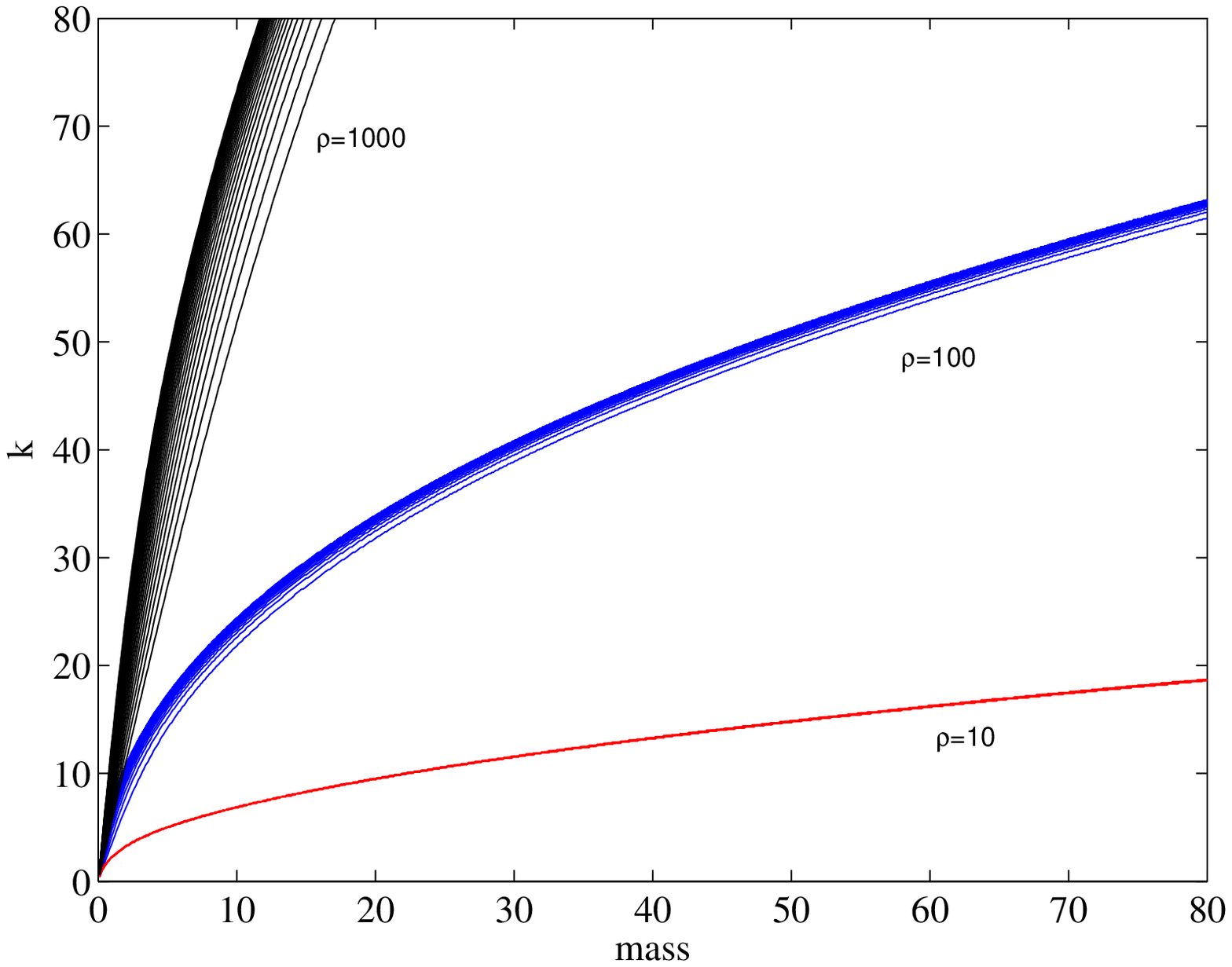}
\end{center}
\caption{(Color online) Optimal $|\bm k|$ as a function of the mass of the field for different vlues of $\epsilon=6,9,12,\dots,99$ and $\rho=10,100,1000$ showing rapid saturation in $\epsilon$. For higher $\epsilon$ this curves are completely insensitive to $\epsilon$ variations, being very little sensitive for smaller values $\epsilon<10$}
\label{Figsat}
\end{figure}

The relationship between mass and the optimal frequency is very sensitive to variations in $\rho$, presenting no saturation. Conversely figure \ref{Figsat} shows that the optimal $|\bm k|$ curve is almost completely insensitive to $\epsilon$. 
All the different $\epsilon$ curves are very close to each other. We can take advantage of this to estimate the rapidity independently of the value of $\epsilon$ using the entanglement induced by the expansion on fermionic fields.

\subsection{Optimal $|k|$ tuning method}

\subsubsection{Part I: Rapidity estimation protocol}

Given a field of fixed mass, we obtain the entanglement for different modes $k_1,\dots,k_n$ of the field. Then the mode $k_i$ that returns the maximum entropy will codify information about the rapidity $\rho$, as seen in figure \ref{Fig8}.  One advantage of this method is that there is no need to assume a fixed $\epsilon$ to estimate $\rho$, since the tuning curves (fig. \ref{Fig8}) are have low sensitivity to $\epsilon$ (fig. \ref{Figsat}). Furthermore this method does not saturate for  higher values of $\rho$ since we can use heavier fields to overcome the saturation observed in figure \ref{peaked}. While one might expect  that heavier masses would mean smaller  maximum entropy, figure \ref{Fig10} shows  that if $\rho$ is  high enough
to force us to look at heavier fields to improve its estimation, the  amount of entanglement will also be high enough due to the scaling properties of $S_E(|\bm k|,m)$. We can therefore safely use more massive fields to do estimate $\rho$ since they better codify its value.  

Hence  we have a method for extracting information about $\rho$ that is not affected by the value of $\epsilon$.  Information about $\rho$ is quite clearly encoded in the optimal $|\bm k|$ curve, which is a direct consequence of the peaked behaviour of $S_E(|\bm k|,m)$.

\subsubsection{Part II: Lower bound for $\epsilon$ via optimal $|k|$ tuning}

\begin{figure}[h]
\begin{center}
\includegraphics[width=.50\textwidth]{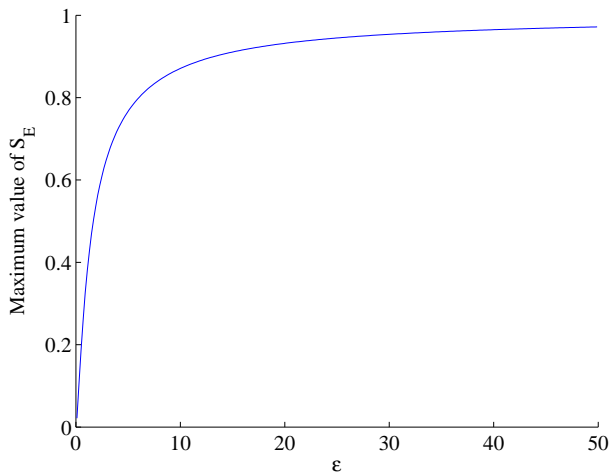}
\end{center}
\caption{$S_E^{\text{max}}(\epsilon)$: Maximum entanglement achievable (optimal $m$ and $|\bm k|$) as a function of $\epsilon$. It does not depend at all on $\rho$}
\label{epsmax}
\end{figure}

We can see from figure \ref{Fig8} that for different values of  $\rho$  the maximum value for the entanglement at the optimal point (optimal $k$ and optimal $m$)  is always $S_E^{\text{max}}\approx0.35$. Consider now $\epsilon\neq 1$. In figure \ref{epsmax} we can see how the maximum entanglement that can be achieved for optimal frequency and mass varies with the volume parameter $\epsilon$. Indeed, the maximum possible entanglement that the optimal mode can achieve is a function of only $\epsilon$ and is independent of $\rho$.
Hence  information about $\epsilon$ is encoded in the maximum achievable fermionic entanglement. Consequently we can find a method for obtaining a lower bound for the total volume of the expansion of the space-time regardless of the value of the rapidity. 

In this fashion we obtain a lower bound for $\epsilon$ since the entanglement measured for the optimal mode is never larger than  the maximum achievable entanglement represented in figure \ref{epsmax}, $S_E(|\bm k|,m) \le S_E^{\text{max}}$. For  instance if the entanglement in the optimal mode is $S_E>0.35$ this will tell us that $\epsilon>1$, whereas if $S_E>0.87$ then we can infer that $\epsilon>10$. Note that as $\epsilon$ increases the entanglement in the optimal $|k|$ mode for the optimal mass field  approaches that of a maximally entangled state when $\epsilon\rightarrow\infty$ .

Although this method presents saturation when $\epsilon\rightarrow\infty$ (being most  effective for $\epsilon \le 20$) its insensitivity to $\bm{\rho}$   means that the optimal $|\bm k|$ method gives us two independent methods for estimating $\rho$ and $\epsilon$. In other words, all the information about the parameters of the expansion (both volume and rapidity) is encoded in the entanglement for the optimal frequency $|\bm k|$.

\subsection{Interpretation for the dependence of $S_E$ on $|{\bm k}|$}

We have seen (figure \ref{bosons}) that for bosons a monotonically decreasing entanglement is observed as $|\bm k|$ increases. By contrast, in the fermionic case we see that there are privileged $|\bm k|$ for which entanglement creation is maximum. These modes are far more prone  to entanglement than any others. 

To interpret  this  we can regard the optimal value of $|\bm k|$ as being associated with a characteristic wavelength (proportional to $|\bm k|^{-1}$) that is increasingly  correlated with the characteristic length of the universe. 
As $\rho$  increases the peak of the entanglement entropy shifts towards higher $|\bm k|$, with smaller characteristic lengths. Intuitively,
 fermion modes with  higher characteristic lengths are less sensitive to the underlying space-time because the exclusion principle impedes the excitation of `very long' modes (those whose $|\bm k|\rightarrow0$).

What about small $|\bm k|$ modes?
As shown in \cite{ball} and in figure \ref{bosons}, for bosons the entanglement generation is higher when $|\bm k|\rightarrow0$. This makes sense because modes of smaller $|\bm k|$ are more easily excited as the space-time expands (it is energetically much `cheaper'  to excite smaller $|\bm k|$ modes).
For fermions entanglement generation, somewhat counterintuitively, decreases for $|\bm k|\rightarrow 0$.   However if 
we naively think of fermionic and bosonic excitations in a box we can appreciate the distinction. We can put an infinite number of bosons with the same quantum numbers into the box. Conversely, we cannot put an infinite number of fermions in the box due to the Pauli exclusion principle. This `degeneracy pressure'  impedes those `very long' modes (of small $|\bm k|$) from being entangled by the underlying structure of the space-time.

\section{Conclusions}\label{conclusions}

We have shown that the expansion of the universe (in a model 2-dimensional setting) 
  generates in entanglement in quantum fields that is qualitatively different
for fermions and bosons.  This result is commensurate with previous studies demonstrating significant differences between the entanglement of bosonic and fermionic fields  \cite{Alicefalls,AlsingSchul,Edu2,Adeschul}.

 We find that the entanglement generated by the expansion of the universe as a function of the frequency of the mode in the fermionic case peaks, while in the bosonic case it monotonically decreases. 
For bosons the most sensitive modes are those  whose $|\bm k|$ is close to zero. However for fermions modes of low $|\bm k|$ are insensitive to the underlying metric. There is an optimal value of $|\bm k|$ that is most prone to expansion-generated entanglement.  This feature may be a consequence of the Pauli exclusion principle, though we have no definitive proof of this.
 
 We have also demonstrated that information about the spacetime expansion parameters  is encoded
  in the entanglement between fermionic particle and antiparticle modes of opposite momenta.  This can be extracted from
the peaked behaviour of the entanglement shown in figure \ref{peaked}, a feature  absent in the bosonic case.  
Information about the rapidity of the expansion ($\rho$) is codified in the frequency of the maximally entangled mode, whereas the information about the volume of the expansion ($\epsilon$) is codified in the amount of entanglement generated for this optimal mode. As   $\epsilon$ tends to infinity the maximum possible $S_E^{\text{max}}$ in the optimal mode approaches the maximally entangled state.

Hence the expansion parameters of spacetime are better estimated from cosmologically generated fermionic entanglement.   Furthermore, these results   show that fermionic entanglement is affected by the underlying spacetime structure in a very counterintuitive way and in a radically different manner than in the bosonic case.  The manner and extent to which these results carry over to $d$-dimensional spacetime remains a subject for future study.

\section{Acknowledgments}

The authors would like to thank an anonymous referee for his/her helpful review and comments on the condensed matter interest of this work.
I. F was supported by EPSRC [CAF Grant EP/G00496X/2] and the Alexander von Humboldt
Foundation and would like to thank Tobias Brandes and his group at
TU-Berlin for their hospitality.  This work was supported in part by the Natural Sciences and Engineering
Research Council of Canada.
E. M-M was supported by a CSIC JAE-PREDOC2007 Grant and by the Spanish MICINN Project FIS2008-05705/FIS.

\bibliographystyle{apsrev}

\end{document}